\documentclass[twocolumn]{aastex61}

\usepackage{lineno}
\usepackage{xspace}
\usepackage{soul}
\usepackage{color}
\usepackage{natbib}
\usepackage{csquotes}
\usepackage{longtable}

\newcommand{\Epeak}{$E_{\rm peak}$\xspace}

\shorttitle{Extragalactic MGFs as SGRBs}

\begin{document}
\AuthorCallLimit=999
\title{Identification of a Local Sample of Gamma-Ray Bursts Consistent with a Magnetar Giant Flare Origin}

\author{E. Burns}
\affiliation{Department of Physics \& Astronomy, Louisiana State University, Baton Rouge, LA 70803, USA}

\author{D. Svinkin}
\affiliation{Ioffe Physical-Technical Institute, Politekhnicheskaya 26, St. Petersburg, 194021, Russia}

\author{K. Hurley}
\affiliation{Space Sciences Laboratory, University of California, 7 Gauss Way, Berkeley, CA 94720-7450, USA}

\author{Z. Wadiasingh}
\affiliation{NASA Goddard Space Flight Center, 8800 Greenbelt Road, Greenbelt, MD 20771, USA}
\affiliation{Universities Space Research Association Columbia, Maryland 21046, USA}

\author{M. Negro}
\affiliation{University of Maryland, Baltimore County, 1000 Hilltop Circle, Baltimore, MD 21250, USA}

\author{G. Younes}
\affiliation{Department of Physics, The George Washington University, Washington, DC 20052, USA}
\affiliation{Astronomy, Physics and Statistics Institute of Sciences (APSIS), The George Washington University, Washington, DC 20052, USA}

\author{R. Hamburg}
\affiliation{Department of Space Science, University of Alabama in Huntsville, Huntsville, AL 35899, USA}

\author{A. Ridnaia}
\affiliation{Ioffe Physical-Technical Institute, Politekhnicheskaya 26, St. Petersburg, 194021, Russia}

\author{D. Cook}
\affiliation{IPAC/Caltech , 1200 E California Blvd, Pasadena, CA 91125, USA}

\author{S. B. Cenko}
\affiliation{NASA Goddard Space Flight Center, 8800 Greenbelt Road, Greenbelt, MD 20771, USA}
\affiliation{Joint Space-Science Institute, University of Maryland, College Park, MD 20742, USA.}

\author{R. Aloisi}
\affiliation{University of Wisconsin-Milwaukee, P.O. Box 413, Milwaukee, WI 53201, USA}
\affiliation{Department of Astronomy, University of Wisconsin-Madison, 475 North Charter Street, Madison, WI 53706, USA}

\author{G. Ashton}
\affiliation{OzGrav: The ARC Centre of Excellence for Gravitational Wave Discovery, Clayton VIC 3800, Australia}

\author{M. Baring}
\affiliation{Department of Physics and Astronomy, Rice University, MS-108, P.O. Box 1892, Houston, TX 77251, USA}

\author{M. S. Briggs}
\affiliation{Department of Space Science, University of Alabama in Huntsville, Huntsville, AL 35899, USA}

\author{N. Christensen}
\affiliation{Artemis, Universit\'e C\^ote d’Azur, Observatoire de la C\^ote d’Azur, CNRS, Nice 06300, France}

\author{D. Frederiks}
\affiliation{Ioffe Physical-Technical Institute, Politekhnicheskaya 26, St. Petersburg, 194021, Russia}

\author{A. Goldstein}
\affiliation{Science and Technology Institute, Universities Space Research Association, Huntsville, AL 35805, USA}

\author{C. M. Hui}
\affiliation{Astrophysics Office, ST12, NASA/Marshall Space Flight Center, Huntsville, AL 35812, USA}

\author{D. L. Kaplan}
\affiliation{University of Wisconsin-Milwaukee, P.O. Box 413, Milwaukee, WI 53201, USA}

\author{M. M. Kasliwal}
\affiliation{Division of Physics, Mathematics, and Astronomy, California Institute of Technology, Pasadena, CA 91125, USA}

\author{D. Kocevski}
\affiliation{Astrophysics Office, ST12, NASA/Marshall Space Flight Center, Huntsville, AL 35812, USA}

\author{O. J. Roberts}
\affiliation{Science and Technology Institute, Universities Space Research Association, Huntsville, AL 35805, USA}

\author{V. Savchenko}
\affiliation{Department of Astronomy, University of Geneva, Ch. d’Ecogia 16, 1290, Versoix, Switzerland}

\author{A. Tohuvavohu}
\affiliation{Department of Astronomy \& Astrophysics, University of Toronto, 50 St. George Street, Toronto, Ontario, M5S 3H4 Canada}

\author{P. Veres}
\affiliation{Department of Space Science, University of Alabama in Huntsville, Huntsville, AL 35899, USA}

\author{C. A. Wilson-Hodge}
\affiliation{Astrophysics Office, ST12, NASA/Marshall Space Flight Center, Huntsville, AL 35812, USA}

\begin{abstract}
Cosmological Gamma-Ray Bursts (GRBs) are known to arise from distinct progenitor channels: short GRBs mostly from neutron star mergers and long GRBs from a rare type of core-collapse supernova (CCSN) called collapsars. Highly magnetized neutron stars called magnetars also generate energetic, short-duration gamma-ray transients called Magnetar Giant Flares (MGFs). Three have been observed from the Milky Way and its satellite galaxies and they have long been suspected to contribute a third class of extragalactic GRBs. We report the unambiguous identification of a distinct population of 4 local ($<$5 Mpc) short GRBs, adding GRB\,070222 to previously discussed events. While identified solely based on alignment to nearby star-forming galaxies, their rise time and isotropic energy release are independently inconsistent with the larger short GRB population at $>$99.9\% confidence. These properties, the host galaxies, and non-detection in gravitational waves all point to an extragalactic MGF origin. Despite the small sample, the inferred volumetric rates for events above $4\times10^{44}$\,erg of $R_{MGF}=3.8_{-3.1}^{+4.0}\times10^5$\,Gpc$^{-3}$\,yr$^{-1}$ place MGFs as the dominant gamma-ray transient detected from extragalactic sources. As previously suggested, these rates imply that some magnetars produce multiple MGFs, providing a source of repeating GRBs. The rates and host galaxies favor common CCSN as key progenitors of magnetars.
\end{abstract}

\section{Introduction} \label{sec:intro} 
The history of GRBs and magnetars are intertwined. Short bursts of gamma-rays were recorded by the Vela satellites beginning in 1967 \citep{klebesadel1973observations}, and were given the phenomenological name GRBs. GRB\,790305B was localized by the InterPlanetary Network (IPN) to the Large Magellanic Cloud  \citep{GMF_790305_LMC,evans1980location}. It was unique in being the brightest event seen at Earth, the prompt emission had a long-lasting, exponentially-decaying, periodic tail \citep{barat1979evidence} and additional, weaker bursts were localized to the same source  \citep{GMF_790305_LMC}. Immediately there were papers investigating if the main event shared a common origin with other GRBs \citep{GMF_790305_LMC_2,cline1980detection}. It is now known to be the first signal identified from a magnetar.

Key results on the nature of GRBs in the subsequent decades were often proven by population-level statistical analysis before direct \enquote{smoking-gun} proof. Perhaps the greatest debate was whether these events had a galactic or an extragalactic origin, with the latter initially disfavored as it would require intrinsic energetics beyond anything previously known. Proof came first indirectly via statistical studies on the spatial distribution of GRBs \citep{meegan1992spatial} and then directly from redshift measurements \citep{metzger1997spectral}.

Studies of the prompt GRB emission provided strong evidence in favor of two populations \citep{Kouveliotou1993}, with short and long GRBs traditionally separated at 2\,s as measured by the $T_{90}$ parameter. Long GRBs were tied to broad-line type Ic core-collapse supernovae called collapsars \citep{galama1998unusual}. The \textit{Neil Gehrels Swift Observatory (Swift)} mission enabled successful detections of afterglow from a sample of short GRBs. Circumstantial evidence pointed towards a neutron star merger origin \citep{Eichler1989,fong2015decade} with direct confirmation that some GRBs arise from binary neutron star mergers came with GW170817 and GRB\,170817A\citep{GW170817-GRB170817A}.

Yet another debate on the behavior of GRBs is whether or not the sources repeated. This is best explained using modern parlance. Soft Gamma-ray Repeaters (SGRs) are galactic magnetars named phenomenologically for the weak, recurrent short bursts that first identified them before their physical origin was known. SGR flares are classified as distinct from GRBs, and have recently been tied to radio emission similar to the cosmological Fast Radio Bursts \citep{bochenek2020fast}. The flare on March 5, 1979 and the subsequent similar events GRB\,980827 \citep{GMF_980827_SGR1900_Mazets,GMF_980827_SGR1900_Hurley} and GRB\,041227 \citep{GMF_041227_SGR1806_Palmer,GMF_041227_SGR1806_Frederiks} from magnetars in the Milky Way are referred to as Magnetar Giant Flares (MGFs). The designation for the prompt emission of MGFs often carries the GRB designation, which we use here. GRBs are now not thought to repeat as collapsars and neutron star mergers are cataclysmic events. While several galactic magnetars have been observed to produce multiple SGR flares, none have been observed to produce multiple giant flares (though this is not surprising). The historic debate on potential repeating GRBs was likely confounded by magnetar transients before the separation of SGR flares from GRBs.

We here refer to GRBs 790305B, 980827, and 041227 as the known MGF sample. The detection of three from the Milky Way and its satellite galaxies implies a high intrinsic rate on a per-galaxy or volumetric basis. These events should be detectable to extragalactic distances by GRB monitors such as Konus-\textit{Wind} \citep{aptekar1995konus}, \textit{Swift}-BAT \citep{Barthelmy05}, and \textit{Fermi}-GBM \citep{Meegan2009}. However, at these distances only the immediate bright spike would be detectable and the event should resemble a short GRB \citep{hurley2005exceptionally}. There are two events discussed in previous literature as extragalatic MGF candidates, being GRB\,051103 \citep{ofek2006short,GMF_051103_M81,hurley2010new} and GRB\,070201 \citep{GMF_070201_M31,ofek2008grb}, whose chance alignment coincidence was measured to be $\sim$1\% \citep{GMF_IPN_Dmitry_search}. 

There have been population-level searches for additional events, which identified no additional candidates \citep{popov2006soft,ofek2007soft,GMF_IPN_Dmitry_search}. However, these studies allow us to constrain the fraction of detected short GRBs that have an MGF origin: \citet{ofek2007soft} show that the rate of galactic events requires this to be $>$1\%, while the lack of additional candidates found in several searches constrain the upper bound to be $<$8\% \citep{tikhomirova2010search,GMF_IPN_Dmitry_search,mandhai2018rate}. These studies and tehir conclusions generally assumed that the brightest MGFs could be detectable to tens of Mpc.

Recently, GRB\,200415A was identified as the third and likeliest extragalactic MGF \citep{IPN2020}. In this work, we perform a new population-level search utilizing the largest GRB sample, new galaxy catalogs that are both more complete and provide additional information, and develop a new formalism to determine if we can prove extragalactic MGFs contribute to the observed GRB population. Section\,\ref{sec:search} details the search formalism which identifies four nearby events, identifying an additional extragalactic candidate. The progenitors of our identified sample are investigated in Section\,\ref{sec:progenitor}, the implications of which are discussed in Section\,\ref{sec:implications}. We conclude with discussions in Section\,\ref{sec:conclusions}.

\section{Local GRBs}\label{sec:discovery}
The \enquote{smoking-gun} evidence of an MGF is the long periodic tails which are modulated by the rotation period of the neutron star \citep{hurley1999giant} and also show quasi-periodic oscillations related to the modes of the neutron star itself \citep{Barat1983,strohmayer2005discovery,israel2005discovery,watts2006detection}. However, these signatures are not unambiguously identifiable at extragalactic distances with existing instruments. As such, we follow prior population-level searches and focus on spatial information: if a well-localized short GRB is an MGF it should occur within $\sim$50\,Mpc and be consistent with a cataloged galaxy. We combine existing GRB and galaxy catalogs to build the most complete set of information from existing literature. For each individual burst we quantify our belief that it is an MGF from a known galaxy through comparison of two PDFs, which are discussed below. These PDFs are generated in HEALPix \citep{gorski2005healpix}. The resolution of HEALPix maps is defined by the NSIDE parameter, where the number of total pixels is equal to the square of the NSIDE times twelve. The maps were generated with NSIDE=8192, corresponding to a pixel width of ~$\sim$0.5 arcminutes.

\subsection{The GRB Sample}\label{sec:GRB_sample}
We utilize data from \textit{CGRO}-BATSE \citep{fishman1989batse}, Konus-\textit{Wind} \citep{aptekar1995konus}, \textit{Swift}-BAT \citep{Barthelmy05}, \textit{Fermi}-GBM \citep{Meegan2009}, and additional information from the IPN\footnote{ssl.berkeley.edu/ipn3/index.html}. Triggers from the same events were matched utilizing temporal information for all events and spatial information \citep{ashton2018coincident} when available. The total sample contains more than 11,000 GRBs observed, with $>$1,200 short GRBs using the standard 2\,s cutoff. 

Our burst sample selection requires three things. We consider only short GRBs ($T_{90}<2$\,s) where the $T_{90}$ used is the shortest reported by any triggering instrument. Second, we require the bolometric fluence (1\,keV-10\,MeV) determined from a broadband instrument (Konus, BATSE, or GBM), converting from the instrument-specific ranges as necessary. Intercalibration uncertainties are within $25\%$. For the trigger times, duration, and spectral properties we utilized the latest catalog information \citep{paciesas1999fourth,svinkin2016second,Lien2016,von2020fourth}, updated online catalogs\footnote{\url{http://www.ioffe.ru/LEA/shortGRBs/Current/index.html}}, GCN circulars, and performed dedicated analysis when necessary.

Lastly, we require well-localized GRBs, constructed from all available information. For BATSE localization we utilize the latest catalogs \citep{goldstein2013batse} and apply the largest systematic error \citep{briggs1999error}. \textit{Swift}-BAT positions are taken from the updated \textit{Swift}-BAT Catalog\footnote{\url{https://swift.gsfc.nasa.gov/results/batgrbcat/index.html}} and \textit{Swift}-XRT localizations are utilized when available\footnote{\url{https://swift.gsfc.nasa.gov/archive/grb_table/}}. \textit{Fermi}-GBM localizations are quasi-circular and were generated using the latest methods \citep{goldstein2020evaluation} for all bursts.

KONUS localizations are an ecliptic band which are summarized in the IPN catalogs. The IPN compiles localization information for GRBs, including the timing annuli derived from the relative arrival times of gamma-rays at distant spacecraft. Information used here is from the IPN localizations of Konus short GRBs through 2020 \citep{pal2013interplanetary} and the IPN list kept up to date online\footnote{\url{http://www.ssl.berkeley.edu/ipn3/}}. Additional IPN localizations were compiled for more than 100 additional short GRBs for this work, which were added to the online table. The location information, including systematic error, from the autonomous localizations, timing annuli, and Earth occultation selections are converted to the HEALpix format using the GBM Data Tools\footnote{\url{https://fermi.gsfc.nasa.gov/ssc/data/analysis/gbm/gbm_data_tools/gdt-docs/}}. These independent PDFs are combined into a final PDF referred to as $P^{GRB}$.

The localization threshold is set to a 90\% confidence area $<4.125$\,deg$^2$ when including systematic error. This value is chosen as it is 1/10,000 the area of the sky, is comparable to the sum of the angular size of galaxies (as defined in the following section) within 200\,Mpc, and is between previously used thresholds \citep{GMF_IPN_Dmitry_search}. With the bolometric fluence measure requirement and the removal of bursts with known redshift \citep{Lien2016} beyond the distance where the event may be a detected MGF, we are left with a sample of 250 short GRBs. We do not apply more stringent cuts on spectral or temporal information at this stage as the relevant parameters are not uniformly reported in GRB catalogs.


\subsection{The Galaxy Sample}\label{sec:galaxy_sample}
For the galaxies considered in this work we require the position (RA, Dec, Distance), angular extent (if non-negligible at our spatial resolution; represented here as ellipses), and the current Star Formation Rate (SFR). The z=0 Multiwavelength Galaxy Synthesis (z0MGS) Catalog \citep{2019ApJS..244...24L} combines the ultraviolet observations from GALEX \citep{morrissey2007calibration} with the infrared observations of WISE \citep{wright2010wide} to uniformly measure gas and dust for galaxies within approximately 50\,Mpc. As a result, for galaxies contained in this catalog these measures of the distance and SFR are our default values. The angular size of galaxies is represented as an ellipse when data allows or as a circle when the axial ratio is not known. Angular extent is taken from the input catalogs, but is generally the Holmberg isophote, i.e. where the B band brightness is 26.5 mag arcsecond$^2$.

The Census of the Local Universe (CLU) Catalog \citep{cook2019census} aims to provide the most complete catalog of galaxies out to 200\,Mpc. We use the CLU measures of distance and SFR when they are not provided by z0MGS, and we use the CLU measures for angular size (which are not provided by the z0MGS). When missing, we add position angle information from HyperLEDA \citep{paturel2003hyperleda}. The SFR measures of these two catalogs correct for internal extinction using WISE4/FUV luminosities. To ensure completeness within $<$10\,Mpc we supplement these two catalogs with the Local Volume Galaxy (LVG) Catalog \citep{karachentsev2013star}. The three catalogs are matched by name, with help from the NASA/IPAC Extragalactic Database (NED)\footnote{\url{https://ned.ipac.caltech.edu/}}, and position information. 

We consider galaxies between 0.5\,Mpc (excluding the Milky Way and its satellite galaxies) and 200\,Mpc (beyond where MGFs can be detected), which leaves more than 100,000 galaxies. The SFR is a key parameter in our method and our inferences also rely on scaling the properties of our host galaxy. The Milky Way SFR used here is 1.65$\pm$0.19 M\,$_\odot/yr$ \citep{licquia2015improved}. We specify the SFR for NGC\,3256, which was identified in \citet{popov2006soft} as being a likely source of detectable extragalactic MGFs. We searched the literature for values of the active SFR in this galaxy and take the value of $\sim$36\,$_\odot/yr$ from \citet{lehmer20150} which is inferred using UV information and is among the middle reported values.

\subsection{MGF Spatial Distribution}\label{sec:PMGF}
We seek an all-sky PDF, $P^{MGF}$, representing the probability that a given position is to produce a MGF with a particular fluence at Earth. Note that this is determined by the fluence of each burst considered, but is constructed independently of the location of the burst itself, $P^{GRB}$. The comparison of the two PDFs generated for each burst quantifies the likelihood that a given short GRB has an MGF origin, which is performed in the next section. This section details the burst-specific construction of $P^{MGF}$.

If a given burst has an MGF origin it should arise from a cataloged galaxy and its intrinsic energetics should fall into the expected range. To construct this we compute a weight for each galaxy representing how likely it is to have produced the observed fluence for the burst under consideration. This weight has two-components: a linear weighting with SFR and a more complex weighting that compares the inferred intrinsic energetics (determined by the burst fluence and potential host galaxy distance) against an assumed PDF.

Magnetars are expected to be able to produce MGFs only for a short period of time \citep[approximately 10 kyr][]{beniamini2019formation}, tying the predicted rate of MGFs to the rate of their formation. The rate of CCSN can be inferred from the SFR since the lifetimes of stars that undergo core-collapse is much shorter than the timescale probed by the SFR tracers \citep{botticella2012comparison}. Under the assumption that the dominant formation channel for magnetars is CCSN (which is explored in Section\,\ref{sec:implications}) we can infer the rate of MGFs from a galaxy from its SFR. Thus, each galaxy is linearly weighted with SFR. We use the far ultraviolet measure of SFR \citep{lee2010galex} when available as it should track massive stars likely to undergo core-collapse, otherwise we use the H$\alpha$ measure \citep{kennicutt1998star} scaled by the average difference from galaxies with both measures to account for the lack of dust correction in the LVG catalog.

Next we can determine the total isotropic-equivalent energetics of a potential burst-galaxy pair as $E_{\rm iso} = 4 \pi d^2 S$ where $S$ is the burst fluence and $d$ the distance to the potential host. This value can be compared to an assumed intrinsic energetics PDF to determine how likely the event is to be an MGF. For example, a particularly high fluence short GRB spatially aligned with a distant galaxy would require an intrinsic energetics far beyond what has been observed in the galactic MGFs, excluding an MGF origin. We note that some studies utilize the peak luminosity $L_{\rm iso}^{Max}$ but we work with an $E_{\rm iso}$ distribution as there is stronger theoretical guidance on the maximum total energy that can be released (related to the magnetic fields of the magnetar) than on the timescale that it is released.

\begin{figure}
\includegraphics[width=\columnwidth]{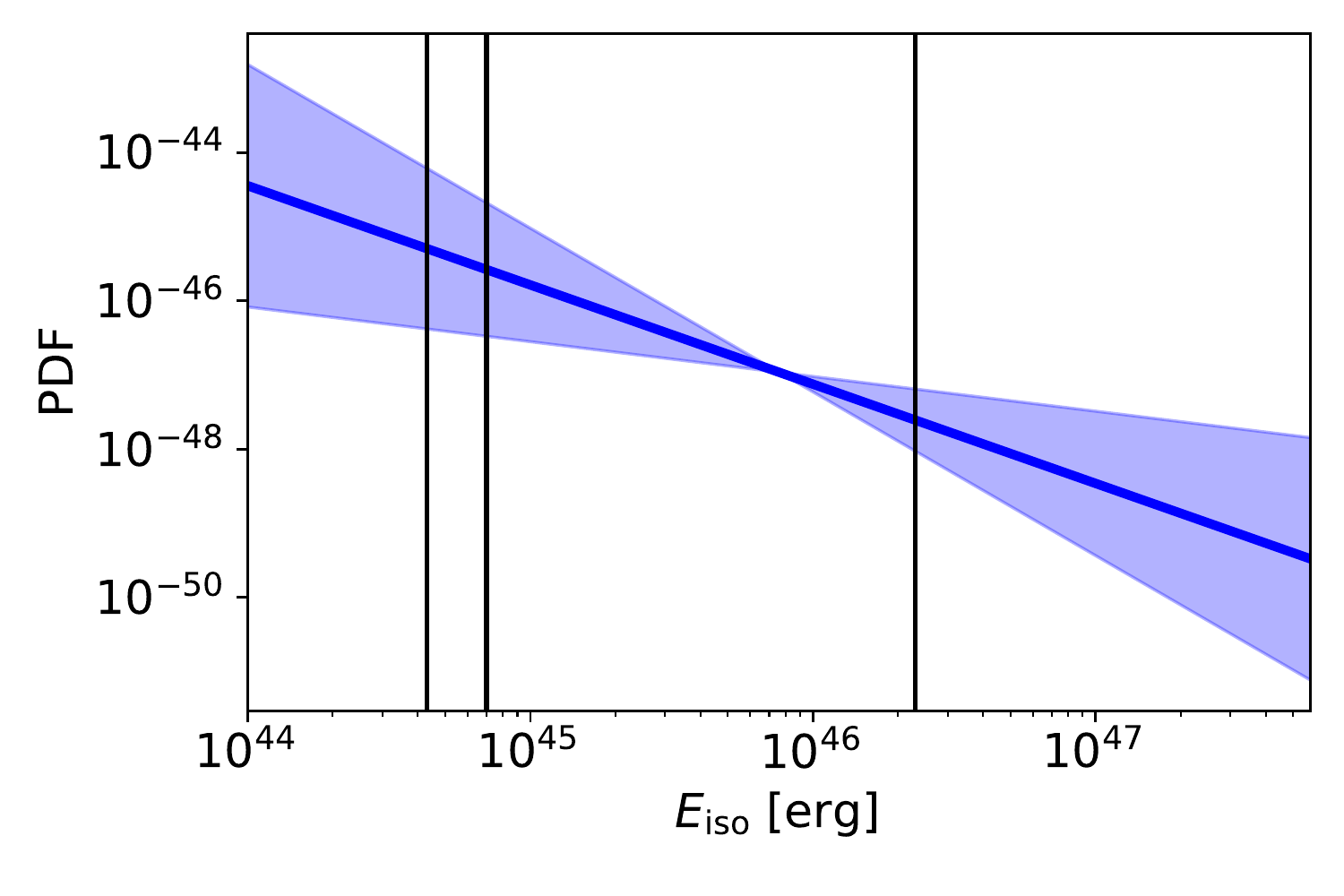}
\caption{The initial assumed MGF energetics distribution, with $E_{\rm iso, min}$ and $E_{\rm iso, max}$ set to the x-axis boundaries. The PDF form is $(1-\alpha)E_{\rm iso}^{-\alpha}/(E_{\rm Max}^{1-\alpha}-E_{\rm Min}^{1-\alpha})$. As described in the text, $\alpha=1.3 \pm 0.9$ (at 90\% confidence). The three $E_{\rm iso}$ values from the known MGFs used to constrain the slope are shown as black vertical lines.}
\label{fig:eMGF_Eiso}
\end{figure}

We now construct an informed intrinsic energetics function, assuming a power-law distribution with an assumed minimum and maximum value, which is similar to the behavior of lower energy magnetar flares \citep{cheng1996earthquake}. Our method bypasses the need for an assumed detection threshold, which is difficult to quantify when considering many instruments over 30 years. The assumed and inferred values are reported below, with the initially determined distribution shown in Figure\,\ref{fig:eMGF_Eiso}.

The slope of a power-law can be determined via maximum likelihood, independent of an assumed maximum value, as \begin{equation}
    \alpha = 1 + n\left[ \sum_{i=1}^n {\rm ln} \left( \frac{E_{\rm iso, i}}{E_{\rm iso,min}} \right) \right]^{-1},
    \sigma_\alpha = \frac{\alpha-1}{\sqrt{n}} + O(n^{-1})
\end{equation}
\noindent where the sum is over the observed $E_{\rm iso}$ and $E_{\rm iso,min}$ is the lowest considered value \cite{newman2005power,bauke2007parameter}. We set $E_{\rm iso,min}$ as $1.0\times10^{44}$\,erg which is a factor of a few below the lowest value measured in a known MGF as shown in Table\,1 but above the brightest SGR flare that lacked the periodic tail emission \citep{GMF_980618_SGR1627_Mazets}. Iterating over the $E_{\rm iso}$ values of the known MGFs (GRBs 790305B, 090827, and 041227) gives $\alpha = 1.3 \pm 0.9$ at 90\% confidence, where we have included the $O(n^{-1})$ error contribution. In order to minimize the required computation we assume the centroid ($\alpha=1.3$) in what follows; the effect of this assumption on our results is discussed in the closing paragraph of this section.



There must be a physical maximum energy for an MGF, which should be related to the total magnetic energy. This is supported by the lack of detections of more energetic events otherwise consistent with an MGF origin. The highest $E_{\rm iso}$ observed for a known MGF is $2.3\times10^{46}$\,erg which comes from the magnetar with the highest reported magnetic field at the surface of $2.0\times10^{15}$\,G \citep{olausen2014mcgill}. We note this reported value is approximately 3 times larger than the dipolar spin-down inferred magnetic field value of 7$\times$10$^{14}$\,G \citep{younes2017sleeping}, but we have confirmed this does not affect our results. To determine an $E_{\rm iso, max}$ for our search we assume a dipole field, where the available energy scales as $B^2$, and a nominal maximum magnetic field strength of $\sim$1.0$\times10^{16}$\,G. This gives $E_{\rm iso, max}=2.3\times10^{46}\,\rm{erg} \times (1.0\times10^{16} G/2.0\times10^{15} G)^2=5.75\times10^{47}$\,erg.

This allows us to determine the burst-specific two-component weight for each of the $>$100,000 galaxies in our sample, which are weighted linearly by its SFR multiplied by the value of the $E_{\rm iso}$ PDF for the inferred energetics considering the burst fluence and galaxy distance. The sum of the galaxy weights is normalized to unity. Then, $P^{MGF}$ is built by placing the calculated weights at the position of the host galaxy. If the angular diameter of the galaxy is larger than the effective resolution of our discrete sky representation ($\sim$arcminute$^2$) then its weight is uniformly distributed over its angular extent. 

\subsection{The Search}\label{sec:search}
For each of the 250 short GRBs in our sample we generate $P^{GRB}$ from the observations of the GRB and $P^{MGF}$ from theoretically motivated expectations. We quantify the likelihood that a given GRB has an MGF origin using $\Omega = 4\pi\sum_i P^{GRB}_i P^{MGF}_i /A_i$ where $P^{GRB}_i$ and $P^{MGF}_i$ indicate the probability for each PDF in the $i$th sky region, which has area $A_i$ \citep{ashton2018coincident}.

\begin{figure}
  \centering
    \includegraphics[width=\columnwidth]{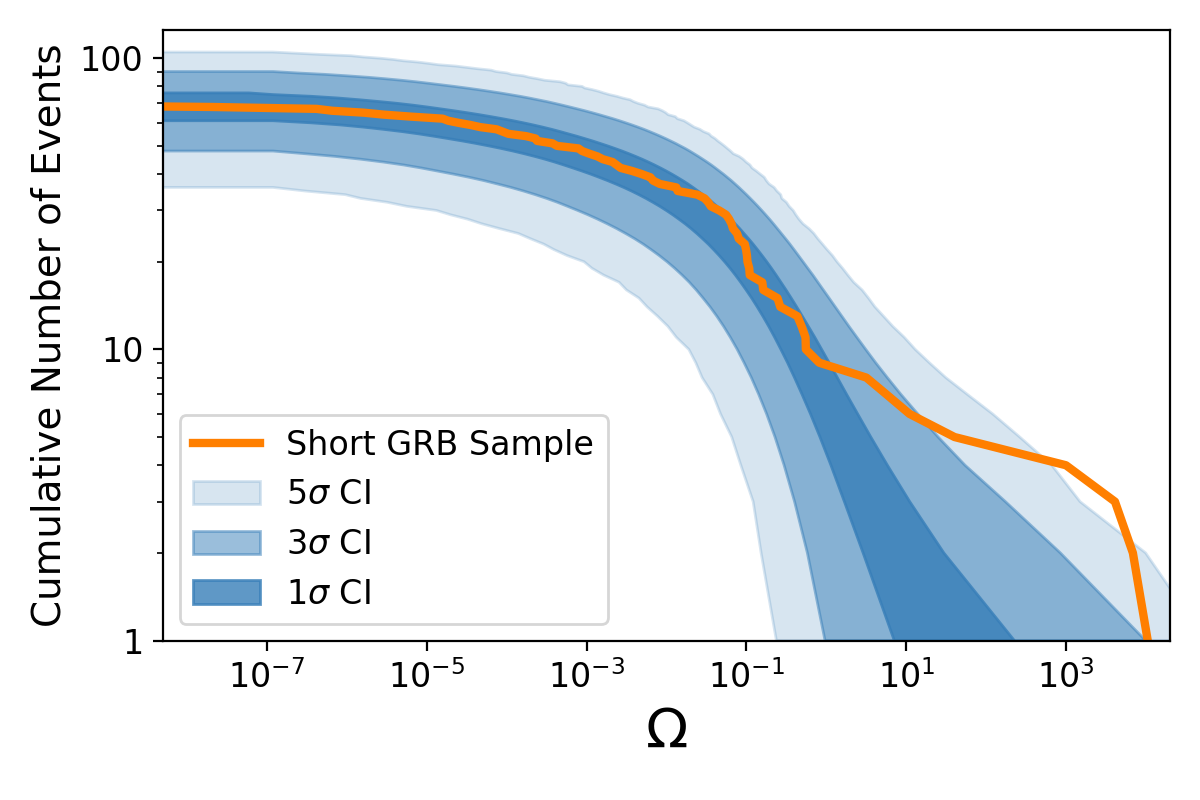}
  \caption{The discovery of a local but extragalactic population of GRBs. $\Omega$ is a statistic that ranks how believable the event is to be an extragalactic MGF, with values for the true population is shown in orange. The background confidence intervals at 1, 3, and 5$\sigma$ are shown in blue. The four most significant events together surpass 5$\sigma$ discovery significance.}
  \label{fig:FAR}
\end{figure}

Significance is determined by the empirical False Alarm method \citep[e.g][]{messick2017analysis} with $\Omega$ as our ranking statistic. Our backgrounds are generated by simulating different galaxy distributions. Each iteration is generated by uniform rotation of the 2D (RA, Dec) positions of the galaxies in our sample, which maintains the distance and SFR distributions as well as local structure. Population-level confidence intervals created through comparison of each rotation against our full GRB sample with results are shown in Figure\,\ref{fig:FAR}. At 3 and 4 events the short GRB sample has an excess surpassing 5$\sigma$ discovery significance, with individual significance values of the four bursts between 1.2$\times10^{-4}$ and $4.9\times10^{-6}$ as given in Table\,\ref{tab:MGFs}.

\begin{figure*}[t]
  \centering
  \includegraphics[width=\textwidth]{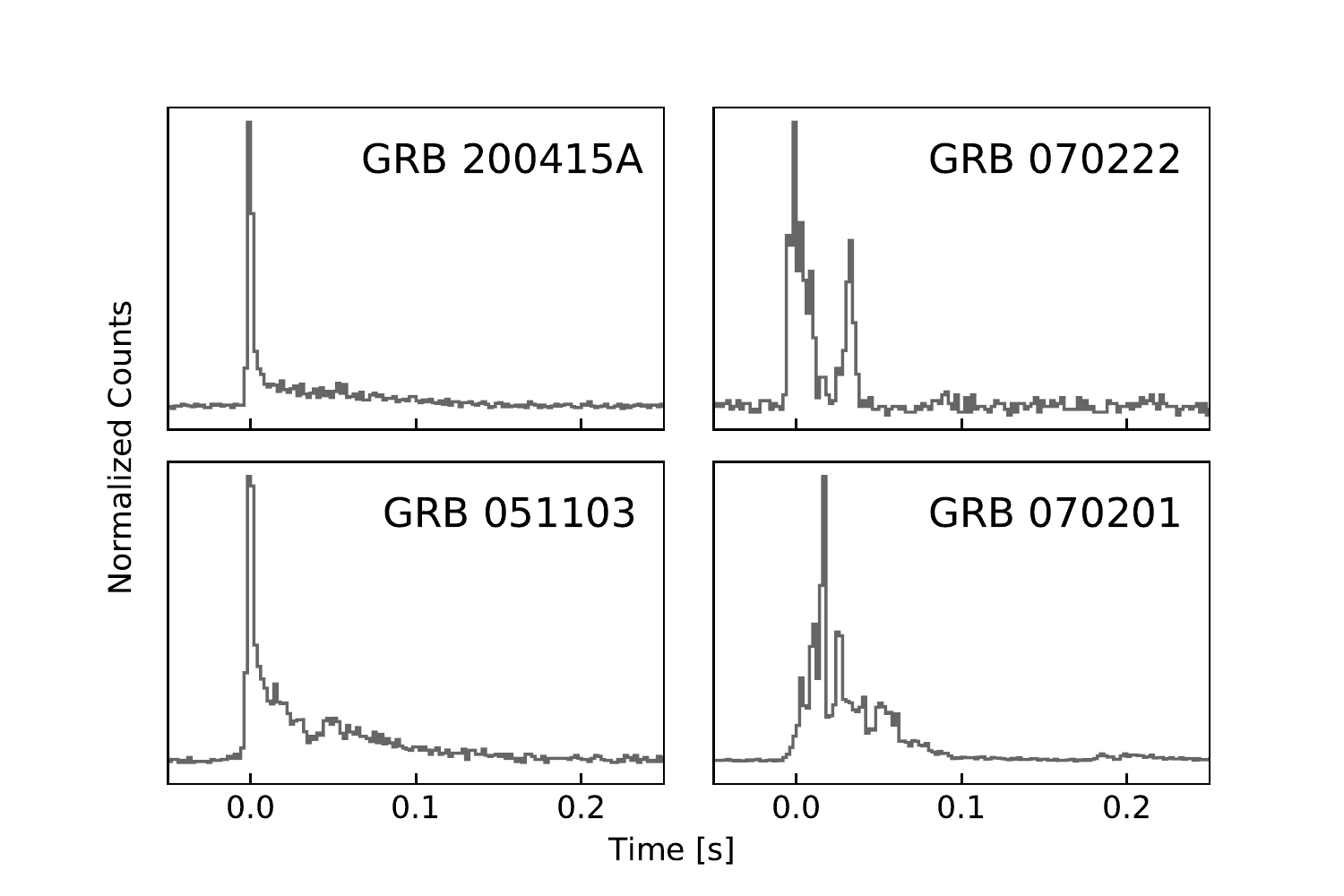}
  \caption{The lightcurves of the candidate extragalactic MGFs in order of significance from Extended Data Table 1. These are from Konus-\textit{Wind} and plotted with 2\,ms resolution \citep{GMF_051103_M81,GMF_070201_M31,IPN2020}, with GRB\,070222 reported here for the first time. While GRBs 200415A and 051103 are strikingly similar \citep{IPN2020} and GRB\,070201 is broadly consistent with a single emission episode, GRB\,070222 has two temporally and spectrally distinct pulses (see Appendix B), suggesting varied behavior.
  }
  \label{fig:lc}
\end{figure*}

Three of the four are discussed in the literature as extragalatic MGF candidates. The Konus-\textit{Wind} lightcurves are shown in Figure\,\ref{fig:lc}. GRB\,070201 has the least robust association to a nearby galaxy; however, the localization is comparatively large ($\sim$10x the other events) and M31 has the largest angular size of any galaxy in our sample, together lowering $\Omega$ even for real associations. We confirm this by checking GRB\,790305B with the Large Magellanic Cloud \citep{evans1980location,cline1982precise}, which has even larger angular extent than M31, giving $\Omega=500$. 

We perform a number of sanity checks to ensure our assumptions do not significantly affect our results. The search we run assuming our centroid $\alpha=1.3$ value; however, we have confirmed that running the search at the 90\% confidence interval bounds ($\alpha=0.5, 2.2$) identifies the same four bursts as significant outliers and does not identify other candidates. Running the search at greater NSIDE affects our $\Omega$ values by $<$10\%. Rerunning the search where the linear SFR weighting is altered to the stellar mass results in identification of the same galaxies but with generally lower $\Omega$ values. Running with specific SFR returns similar results. Together these suggest a progenitor that tracks SFR. Our results are insensitive to the assumed $E_{\rm iso, min}$, so long as we do not exclude known events, as events of this strength are not detected far into the universe. There are a few events with $\Omega>1$ which are either excluded as events of interest for our MGF search or insignificant given our sample. Lastly, significantly raising the assumed $E_{\rm iso, max}$ marginally identifies GRB\,100216A ($\Omega=10$) which indeed has a potential host galaxy within 200\,Mpc \citep{2010GCN.10429....1P}, which is inconsistent with expectations for MGFs.

\begin{table*}[t]
    \centering
    \begin{tabular}{|r|ccc|cccc|}
\hline													
	&	Known	&		&		&	Extragalactic	&		&		&		\\ \hline
MGF Event	&	790305B	&	980827	&	041227	&	200415A	&	070222	&	051103	&	070201	\\ \hline \hline
Origin	&		&		&		&		&		&		&		\\ \hline
False Alarm Rate	&	0	&	0	&	0	&	$4.9\times10^{-6}$	&	$7.8\times10^{-6}$	&	$1.5\times10^{-5}$	&	$1.2\times10^{-4}$	\\
BNS Excl. [Mpc]	&		&		&		&		&	6.7	&	5.2	&	3.5	\\ \hline \hline
Galaxy Properties	&		&		&		&		&		&		&		\\ \hline
Catalog Name	&	LMC	&	MW	&	MW	&	NGC253	&	M83	&	M82	&	M31	\\
Distance [Mpc]	&	0.054	&	0.0125	&	0.0087	&	3.5	&	4.5	&	3.7	&	0.78	\\
SFR [$M_\odot/yr$]	&	0.56	&	1.65	&	1.65	&	4.9	&	4.2	&	7.1	&	0.4	\\ \hline \hline
GRB Properties	&		&		&		&		&		&		&		\\ \hline
Duration [s]	&	$<$0.25	&	$<$1.0	&	$<$0.2	&	0.100	&	0.038	&	0.138	&	0.010	\\
Rise Time [ms]	&	$\sim$2	&	$\sim$4	&	$\sim$1	&	2	&	4	&	2	&	24	\\
$L_{\rm iso}^{Max}$  [$10^{46}$ erg/s]	&	0.65	&	2.3	&	35	&	140	&	40	&	180	&	12	\\
$E_{\rm iso}$ [$10^{45}$ erg]	&	0.7	&	0.43	&	23	&	13	&	6.2	&	53	&	1.6	\\
Index	&		&		&	-0.7	&	0.0	&	-1.0	&	-0.2	&	-0.6	\\
\Epeak [keV]	&	~500	&	~1200	&	850	&	1080	&	1290	&	2150	&	280	\\
\hline
    \end{tabular}
\caption{A summary of the MGF sample. Significance for extragalactic events is from this text. BNS Excl. refers to the neutron star merger exclusion distances from LIGO. LMC refers to the Large Magellanic Cloud and MW refers to the Milky Way. Individual significance is determined by comparison of the individual $\Omega$ against the full background sample. Distances for the known magnetars come from \citet{olausen2014mcgill}; extragalactic distances are taken from the host galaxy values (which have minor variations with our catalog values). GRB parameters include \Epeak as the energy of peak output, Index is the low-energy power-law from the spectral fit, and the rest are discussed in the text. GRB measures for the galactic events are from the literature; GRB measures for extragalactic events are all measured from Konus-\textit{wind} data.
}
\label{tab:MGFs}
\end{table*}

\section{Progenitor Investigations}\label{sec:progenitor}
To determine the origin of these four bursts we first determine if the known GRB progenitors are compatible. Collapsars power long GRBs with durations $\gtrsim$2\,s and are followed immediately by afterglow and then by broad-lined type Ic supernovae. This origin is excluded as all four events have durations 0.1\,s or less. Additionally, no subsequent supernova were reported in any case (\citealt{li2011nearbyii}; though see \citealt{gehrels2006new,grupe2007swift}). A neutron star merger origin is excluded by LIGO non-detections in gravitational waves for three of the four events \citep{abbott2008implications,abadie2012implications,aasi2014search}, but observations are insufficiently sensitive to inform on the origin of GRB\,200415A. One may consider if off-axis GRBs could explain these events. The best known such event is GRB\,170817A where the duration was longer and spectrum softer than the bulk of the short GRB population, which is inconsistent with the prompt emission from these four local events. Further, the rates of these local events (discussed in the following section) are orders of magnitude higher than cosmological GRBs \citep{siegel2019collapsars}, even considering events that are oriented away from Earth. 

\begin{figure}
\begin{minipage}[c]{8.9cm}
\includegraphics[width=\textwidth]{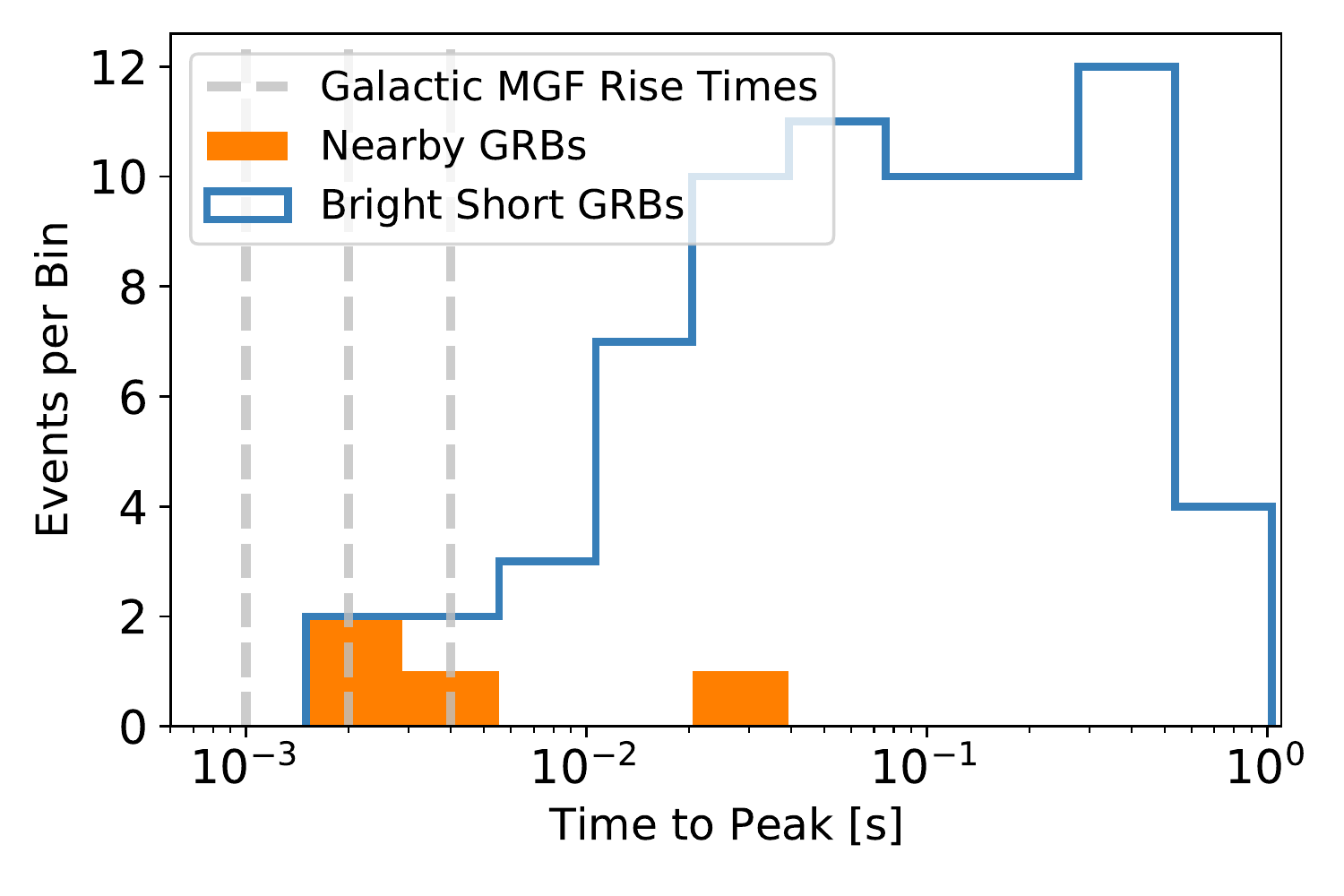}
\end{minipage}
\begin{minipage}[c]{8.9cm}
\includegraphics[width=\textwidth]{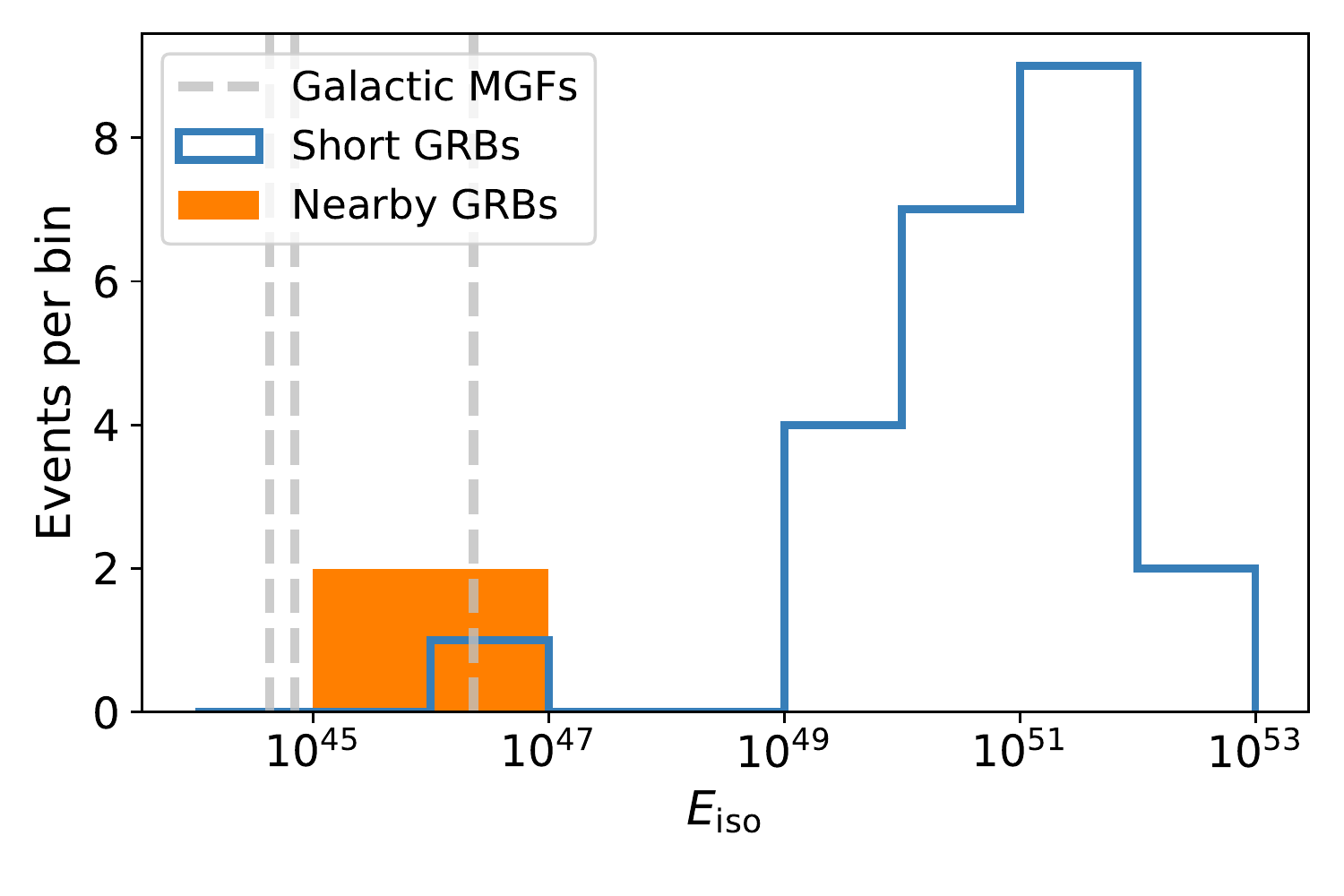}
\end{minipage}
\caption{Key parameter comparison of the extragalactic MGF candidates against the wider short GRB population and the known MGFs. \textit{Top} shows the Time to Peak Konus distributions and \textit{bottom} the $E_{\rm iso}$ distributions. The only comparable $E_{\rm iso}$ value for a burst from a neutron star merger is the off-axis GRB\,170817A.}
\label{fig:sample_comparison}
\end{figure}

To determine the progenitors of these events we follow the historical procedure, where we begin by population comparison of prompt emission parameters. The only additional potential progenitor for extragalatic GRBs commonly discussed in the literature are MGFs where, contrary to the works that identified the two confirmed progenitors, we have the advantage of observations of galactic events which are summarized in Table\,\ref{tab:MGFs}. The parameters relevant for only the main peak of the flare that appear distinct from cosmological GRBs are the rise time and the intrinsic energetics. Figure\,\ref{fig:sample_comparison} contains the population comparison of these parameters.

First, MGFs have rise times of order a few ms, far shorter than most cosmological short GRBs \citep{hakkila2018properties}. Rise times are not reported in most GRB catalogs. As a proxy for the rise time we define the Time to Peak as the time from the start of the emission to the beginning of the peak 2\,ms counts interval. An Anderson-Darling k-sample test against 75 bright Konus short GRBs ($\sim$15\% brightest bursts detected by Konus between 1994 and 2020) rejects the null hypothesis that they are drawn from the same population at $>$99.9\% confidence. 

Second, MGF $E_{\rm iso}$ values are orders of magnitude fainter than cosmological GRBs, where only the unusual GRB\,170817A \citep{GW170817-GRB170817A} is comparable. This parameter depends on the distance to the source, which is not directly observable from prompt emission. For some cosmological GRBs direct distance (redshift) determination is made from follow-up observations. However, for most short GRBs the distance is determined by first robustly associating the short GRB to an aligned or nearly aligned host galaxy, and then determining the distance to the host \citep{fong2015decade}. We adapt this last approach for MGFs to enable the use of larger prompt emission localizations and expected host galaxy properties. For each GRB and potential host galaxy we calculate $\Omega_{Host} = 4\pi\sum_i P^{GRB}_i P^{Host}_i /A_i$ with $P^{Host}$ the weighted spatial distribution of that galaxy. Each GRB has only a single likely host, providing robust association. GRB\,051103 has been discussed in the literature as belonging to the M81 Group of galaxies \citep{GMF_051103_M81}, which is dominating by the interacting galaxies M81 and M82. Our galaxy catalog selection and method assigns the burst to M82.

The inferred $E_{\rm iso}$ values for each extragalatic MGF candidate is given in Table\,\ref{tab:MGFs}. For the population comparison we add the $E_{\rm iso}$ distribution of GBM short GRBs \citep{GW170817-GRB170817A} to the sample of Konus bursts with measured redshift \citep{tsvetkova2017konus}. Together these give 23 short GRBs with $E_{\rm iso}$ determined by a broadband instrument, which is the largest such sample to date. The extragalactic MGFs are clearly inconsistent with the broader population, rejecting the null hypothesis at $>$99.9\% confidence.

Host galaxy studies of GRBs have been key in determining prior progenitor channels \citep[e.g.][]{fong2015decade}. As discussed in the design of our method, MGFs are expected to arise in star-forming galaxies or star-forming regions. Within our maximal detection distance for these bright events the galaxies with the highest SFR are M82, M83, NGC\,253, and NGC\,4945 \citep{mattila2012core}. GRB 051103 is associated to M82 by our method or is consistent with star-forming knots on the outskirts of M81 \citep{ofek2006short}, GRB 070222 to M83, and GRB 200415A to the star-forming core of NGC\,253 \citep{IPN2020}. GRB 790305B is associated to the star-forming Large Magellanic Cloud. This is consistent with a massive-star progenitor, as expected for an MGF origin.

Individually, GRBs\,200415A and 051103 are the most robust identifications of extragalactic MGFs based on our significance assessment and the results of partner analyses including lightcurve morphology and sub-millisecond variation of the prompt emission \citep{IPN2020,GBM2020}. Newly identified is GRB\,070222 which is in-class with key properties of MGFs. However, it has two distinct but overlapping pulses, which is not known to occur from galactic events. This requires either a broader morphology of MGFs, a distinct and unknown origin, or a 1 in 100,000 chance alignment (Table\,\ref{tab:MGFs}). However, given the range of (quasi-)periodic oscillations seen from magnetar emission such a morphology is not necessarily surprising.

To summarize the observational case for an MGF origin: these events localize to the nearby universe and in particular to star-forming regions or star-forming galaxies. The prompt emission is inconsistent with a collapsar origin and gravitational wave observations exclude a compact merger involving neutron stars and/or black holes. The event rates, quantified below, are in excess of the majority of energetic astrophysical transients but are consistent with predictions from the known MGFs. The properties of the prompt emission are distinct from the larger short GRB population but again consistent with the properties from the known MGFs. There is additional evidence for individual events in partner analyses. We conclude that we have confirmed a sample of extragalatic MGFs that match prior predictions on detection rates and properties from both theoretical and observational studies.

A remaining question is: why have we not identified MGFs to greater distances? Previously, MGFs were thought to be detectable to tens of Mpc. The spectra of the initial pulse of GRBs\,200415A, 051103, and GRB\,070222 are particularly spectrally hard with shallow spectral index and high peak energies, which is consistent with GRB\,041227 \citep{GMF_041227_SGR1806_Frederiks}. Assuming a cut-off power-law spectrum for bright MGFs with a low-energy spectral index $\approx0.0$ and peak energies $\approx$1.5\,MeV produces only 15-20\% of the photons in the nominal triggering energy range of 50-300\,keV as compared to a typical short GRB \citep[assuming an index of 0.4 and peak energy 0.6 MeV;][]{GBM_only_paper}. GRB monitors are triggered by photon counts, which suggests that the harder spectrum reduces the detectable by a factor of $\sim$5 and therefore the volume by a factor of more than 100. Instrument-specific comments are given in Appendix A. Further, there is a local overdensity within $\sim$5\,Mpc of CCSN \citep{mattila2012core}, which provides additional explanation of detections within this range and lack of detections beyond it.

\section{Inferences}\label{sec:implications}
We now proceed to make population-level inferences utilizing the three known MGFs and treating all four of our events as extragalatic MGFs.

\subsection{Intrinsic Energetics Distribution}
The power-law distribution of the energetics of normal SGR flares gave hints to the physical process that produces them \citep{cheng1996earthquake}. Thus, it is interesting to measure the slope of the $E_{\rm iso}$ distribution for MGFs. We assign our search volume and detection threshold by empirical means, selecting $2.0\times10^{-6}$\,erg\,cm$^{-2}$ for the IPN and a maximal detection distance of $\sim$5\,Mpc. We further restrict our sample to the past 27 years, where we have sufficient sensitivity to extragalactic events, leaving the 6 most recent bursts (excluding GRB\,790305B). 

We assume the same power-law functional form for the $E_{\rm iso}$ PDF as our search method; however, we cannot utilize the maximum likelihood estimate because it requires the assumption that the observed sample is complete, which is not true for MGFs at extragalactic distances. Instead we simulate a large number of extragalactic MGFs by drawing $E_{\rm iso}$ from PDFs over a range of $\alpha$ values, assigning them to specific host galaxies weighted by their SFR, and setting the event distance as the host galaxy distance. Events that would be detected are those where the sampled $E_{\rm iso}$ and distance produce a flux greater than our detection threshold. $E_{\rm iso, min}=3.7\times10^{44}$\,erg is determined by sampling the Kolmogorov-Smirnov test statistic value over a range of viable options \citep{bauke2007parameter}. Then, we calculate an Anderson-Darling k-sample value for a range of potentially viable $\alpha$ values. We take the 5\% rejection values as the bounds on a 90\% confidence interval, and determine the mean assuming a symmetric Gaussian distribution, giving $\alpha=1.7 \pm 0.4$. We note that this is consistent with the reported slope values of $5/3$ \citep{cheng1996earthquake} and 1.9 \citep{gotz2006two} recurrent flares from galactic SGRs.

\subsection{Rates}
Utilizing the same sample and selection above we can constrain the intrinsic volumetric rate of MGFs. The dominant sources of uncertainty are the Poisson uncertainty and the imprecisely known sample completeness. The latter is limited by the uncertainty on the power-law index of the intrinsic energetics function, where for a steep index the majority of events will be missed (with most events below $1.0\times10^{45}$\,erg missed in our sample volume) and for a shallow index most events are recovered. The $\alpha$ distribution is taken as a Gaussian. The SFR within 5\,Mpc is 35.5\,M$_\odot/yr$ which is scaled to a volumetric rate by considering the total SFR within 50\,Mpc, which is $\sim$4000\,M$_\odot/yr$ from our galaxy sample. We infer a volumetric rate of $R_{MGF}=3.8_{-3.1}^{+4.0}\times10^5$\,Gpc$^{-3}$\,yr$^{-1}$. 


\subsection{Magnetar Formation Channel}

Magnetars may be generated in a variety of events including common CCSN, low-mass mergers \citep{2006Sci...312..719P}, a rare evolution of white dwarfs \citep{2007ApJ...669..585D}, or a rare sub-type of CCSN such as collapsars or superluminous supernovae \citep{nicholl2017magnetar}. Each of these is consistent with the observed association of magnetars to supernova remnants \citep{beniamini2019formation}. Low-mass merger events have long inspiral times and should track total stellar mass rather than the current SFR, which is disfavored given our model preference for SFR over stellar mass and the discovery of the first MGF from the LMC. A CCSN origin would arise from regions with high rates of star formation. This is consistent with our observations and bolstered by both the lack of detections beyond 5\,Mpc due to the local SFR overdensity and the detection of GRB\,790305B from the low-mass, star-forming Large Magellanic Cloud. The host galaxies of our extragalactic sample and the Milky Way itself have larger mass and higher metallicity than is typically seen in hosts of collapsars or superluminous supernovae \citep{2019arXiv191109112T}. Therefore, the types of host galaxies favor common CCSN as the dominant formation channel of magnetars.

Additional support for this conclusion is provided from the event rates. We can relate our inferred MGF rates to progenitor formation rates as $R_{MGF} = R_{Event} f_M \tau_{Active} r_{MGF/M}$ \citep{tendulkar2016radio} where $R_{Event}$ is the rate of events that may form magnetars, $f_M$ is the fraction that successfully form magnetars, $\tau_{Active}$ the timescale that magnetars can produce MGFs, and $r_{MGF/M}$ the rate of MGFs per magnetar. We take $\tau_{Active} \approx 10^4$\,yr limited by the decay of the magnetic field \citep{beniamini2019formation}. Given the incompleteness of our known magnetar sample and lack of understanding which magnetars can produce MGFs, we use only the 3 known to be capable to estimate an upper bound of $r_{MGF/M}<0.02$\,yr$^{-1}$ per magnetar. We note this is significantly weaker than those reported in the literature that consider all known SGRs, being $\sim 1 \times 10^{-4}$\,yr\,SGR \citep[e.g.][]{ofek2007soft,GMF_IPN_Dmitry_search}.

Of the discussed formation channels only CCSN are expected to track star-forming regions and have a comparable rate, being 7$\times10^4$\,Gpc$^{-3}$\,yr$^{-1}$ in the local universe \citep{li2011nearby}. A fiducial value on $f_M$ is 0.4 with a 2$\sigma$ confidence interval of 0.12-1.0 \citep{beniamini2019formation}; other estimates range between 0.01 and 0.1 \citep[e.g.][]{woods2004soft,2015MNRAS.454..615G}. We require either that some magnetars produce multiple MGFs or that both $f_M\approx1$ and the true rate of $R_{MGF}$ is near our 95\% lower bound. Alternatively, using the CCSN rate and the 95\% lower limit on $R_{MGF}$ we can place observational constraints using our results of $f_M>0.005$, further excluding particularly rare sub-types of, and favoring common, CCSN as the dominant formation channel of magnetars.

\section{Conclusions}\label{sec:conclusions}
To summarize our conclusions:
\begin{itemize}
    \item We have shown that 4 short GRBs occurred within $\sim$5\,Mpc which are the closest events by an order of magnitude in distance. Our analysis was the first to identify GRB\,070222 as a local event.
    \item They are inconsistent with a collapsar or neutron star merger origin.
    \item Their prompt emission is inconsistent with the properties of cosmological GRBs, but is consistent with the observations of the known MGFs.
    \item They originate from star-forming regions or star-forming galaxies, including those with metallicity that prevents collapsars from occurring.
    \item Altogether this matches expectations for an MGF origin, which appear to produce 4 out of 250 events. This would be $\sim$2\% of detected short GRBs (consistent with the 1-8\% range from the literature \citealt{ofek2007soft,GMF_IPN_Dmitry_search}) or $\sim$0.3\% of all detected GRBs.
    \item Modeling the intrinsic energetics distribution of MGFs as a power-law constrains the index to be $1.7 \pm 0.4$.
    \item The volumetric rates are $R_{MGF}=3.8_{-3.1}^{+4.0}\times10^5$\,Gpc$^{-3}$\,yr$^{-1}$.
    \item The rates and host galaxies of these events favor CCSN as the dominant formation channel for magnetars, requiring at least 0.5\% of CCSN to produce magnetars.
    \item We estimate the rate of MGFs per magnetar to be $\lesssim 0.02$\,yr$^{-1}$.
    \item Our results suggest that some magnetars produce multiple MGFs: this would be the first known source of repeating GRBs.
    \item GRB\,070222 suggests MGFs can have multiple pulses.
    \item MGFs may not be detectable to tens of Mpc with existing instruments due to their spectral hardness.
\end{itemize}

Our analysis suggests additional extragalactic MGFs may be identified with improved analysis but \enquote{smoking-gun} confirmation likely requires future instruments. The inferred rates are sufficiently high that they may contribute to the stochastic background of gravitational waves. This, and the recent observations of a fast radio burst to lower-energy gamma-ray flares from magnetars \citep{bochenek2020fast,marcote2020repeating,ridnaia2020peculiar,li2020identification}, suggest the coming years will bring new insights into the physics and emission of magnetars. \\

\noindent \textbf{Acknowledgements}

\noindent N. Christensen is supported by the NSF grant PHY-1806990. The \textit{Fermi} GBM Collaboration acknowledges the support of NASA in the United States under grant NNM11AA01A and of DRL in Germany. The CLU galaxy list made use of the NASA/IPAC Extragalactic Database (NED) which is funded by the National Aeronautics and Space Administration and operated by the California Institute of Technology, and was supported by the GROWTH (Global Relay of Observatories Watching Transients Happen) project funded by the National Science Foundation under PIRE Grant No 1545949.

\bibliography{bibliography}

\section{Appendix A}



We present rough estimates for the maximal detection distance of bright MGFs with representative active instruments. Konus-\textit{Wind} can detect bright MGFs to $\sim$13-16\,Mpc, based on GRBs 051103 and 200415A \citep{IPN2020}. This can be taken as the approximate detection distance of the IPN \citep{GMF_IPN_Dmitry_search}. The following investigations assume a hard spectrum based on the time-integrated values for the most energetic bursts, with a low-energy spectral index $\approx0.0$ and peak energies $\approx$1.5\,MeV. This has only 15\% (20\%) the number of photons over the 15-150\,keV (50-300\,keV) energy range, reducing the detection distance by $\sim$x5 and thus a reduction in volume of $>$100. 

The GBM GRB trigger algorithms cover ~50-300 keV where the short GRB sensitivity is usually quoted over the 64\,ms timescale. With the assumed spectral and energetics values GBM would have only triggered these on-board algorithms out to $\sim$15-20 Mpc. At greater distances only the peak flux interval would be visible, which would be spectrally harder, and reduce this distance. GBM localizations alone are insufficient to associate events to any specific burst. Ground-based searches for GRBs and Terrestrial Gamma-ray Flashes should be able to recover additional events, but may require confirmation in other GRB instruments

The INTEGRAL SPI-ACS and IBIS are especially sensitive to hard and short bursts, and additional extragalactic MGFs have likely triggered the SPI-ACS real-time pipeline in the past.  However, SPI-ACS and IBIS lack the capacity to discriminate extragalactic MGFs from high cosmic ray effects which appear similar to real events in these instruments. The real-time IBAS pipeline has not been tuned to favor short and hard events. We estimate that SPI-ACS would record sufficient signal from extragalactic MGFs for association with another instrument up to 25-35 Mpc, but would only independently report much brighter events out to 15-20 Mpc. Sensitivity of IBIS is close or better than to that of SPI-ACS in about 10\% of the sky, and in the majority of directions, IBIS would only yield detectable signal for extragalactic MGF flares out to at most 10~Mpc. However, PICsIT may often be more suitable for triangulation, owing to better time resolution, and can provide some spectral characterization.

The \textit{Swift} BAT has $>$500 different rate trigger criteria running in real-time onboard, continuously sampling and testing trigger timescales from 4ms up to 64 seconds, each of which is evaluated for 36 different combinations of energy ranges and focal plane regions. While the BAT detector  is sensitive to photons with energies up to ~500 keV, the transparency of the lead tiles in the mask above 200 keV limits its imaging energy range (necessary for a successful autonomous trigger) to 15-150 keV. This narrow and low energy range limits the BAT's sensitivity to hard events, such as MGFs, despite its high effective area. Due to the number and complexity of the onboard triggering algorithms, the varying compute load on the BAT CPU, as well as the evolving state of the BAT detector array and changing operational choices for trigger vetoes/thresholds, modelling the likelihood of an onboard autonomous trigger is quite difficult. In addition, due to BAT's high effective area, continuous time-tagged event data cannot be downlinked, making it difficult to assess the relative completeness of the triggering algorithms vs ground searches, though this is partly ameliorated by GUANO \citep{guano}. Under the assumed energetics and spectral values, we estimate that as of 2020 (averaging ~half of the original detector array online) \textit{Swift}/BAT should reliably trigger on MGFs out to $\sim$25 Mpc in the highest coded region of its field of view. Ground analyses in the downlinked BAT event data can extend this, but the availability of this data will often depend on an external trigger (e.g. GUANO). We note that operational changes to the BAT onboard triggering thresholds with the goal of increasing sensitivity to extragalactic MGFs and local low-luminosity GRBs have been previously attempted. In 2012 the threshold for a successful trigger from an image was lowered from the usual value of 6.5 to 5.7, with the condition that triggers in this range be localized to within 12 arcminute projected offset from a local catalogued galaxy stored in the BAT onboard catalog. No local GRB-like source was ever identified in this program.

\section{Appendix B}

\begin{figure}
    \centering
    \includegraphics[width=8.9cm]{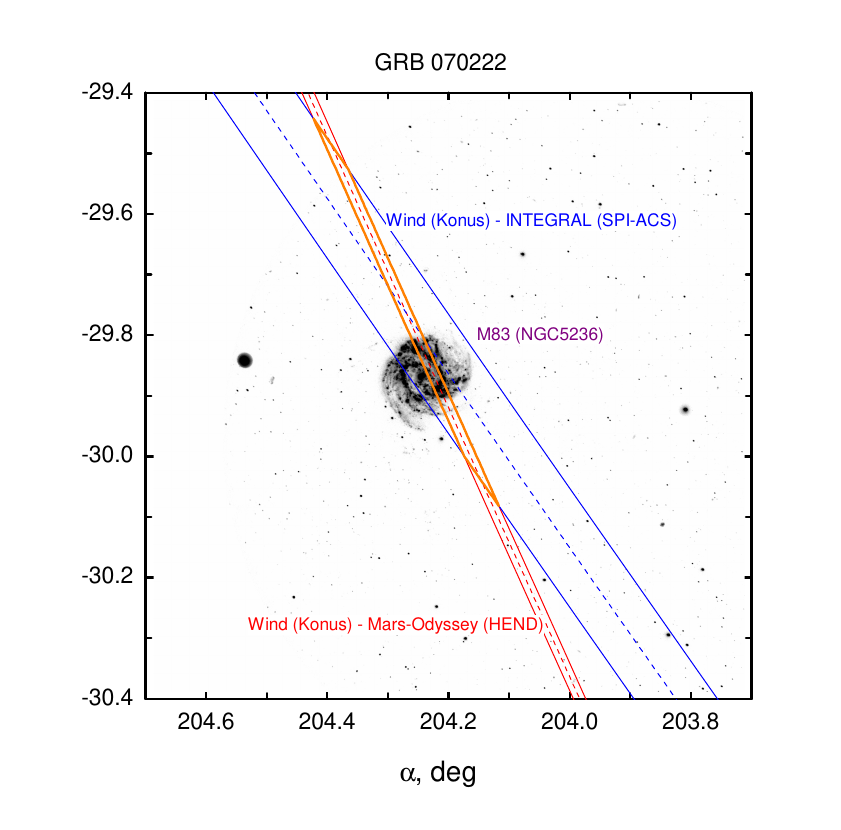}
    \caption{The localization of GRB\,070222 compared to the position and angular size of M83.}
    \label{fig:070222}
\end{figure}

As GRB\,070222 has not been reported elsewhere we describe its basic analysis here. The event was detected by Konus-\textit{Wind}, HEND on Mars Odyssey, and both SPI-ACS and PICsIT on INTEGRAL. Combination of the two best annuli produce a localization with a 90\% containment region of 0.004\,deg$^2$. This location and its consistency with M83 is shown in Figure\,\ref{fig:070222}.

This burst is distinct from the separate candidates as having two separate pulses. Time-resolved analysis of this burst is summarized in Table\,\ref{tab:GRB070222} while time integrated analysis is reported in the Second Konus GRB Catalog \citep{svinkin2016second}.

\begin{table}
    \centering
    \begin{tabular}{|c|c|c|c|c|}
    \hline
    $T_{Start}$ & $T_{Stop}$ & Index & $E_{\rm Peak}$ & Flux \\
    $[s]$ & $[s]$ & & $[keV]$ & $[1\times10^{-6}$\,erg/s/cm$^2]$ \\
    \hline
    -0.006 &   0.012 &  $0.14_{-0.24}^{+0.28}$ &  $733_{-99}^{+138}$    & $153.4_{-16.5}^{+21,2}$ \\
    0.026  & 0.038 & $-0.27_{-0.36}^{+0.48}$ &   $193_{-14}^{+25}$ & $24.5_{-3.0}^{+3.0}$ \\
    \hline
    \end{tabular}
    \caption{The time-resolved analysis of the two pulses of GRB\,070222. Errors are quoted at 68\% confidence. The main pulse is spectrally hard, similar to the time-integrated fits of GRB\,200415A and GRB\,051103.}
    \label{tab:GRB070222}
\end{table}

\end{document}